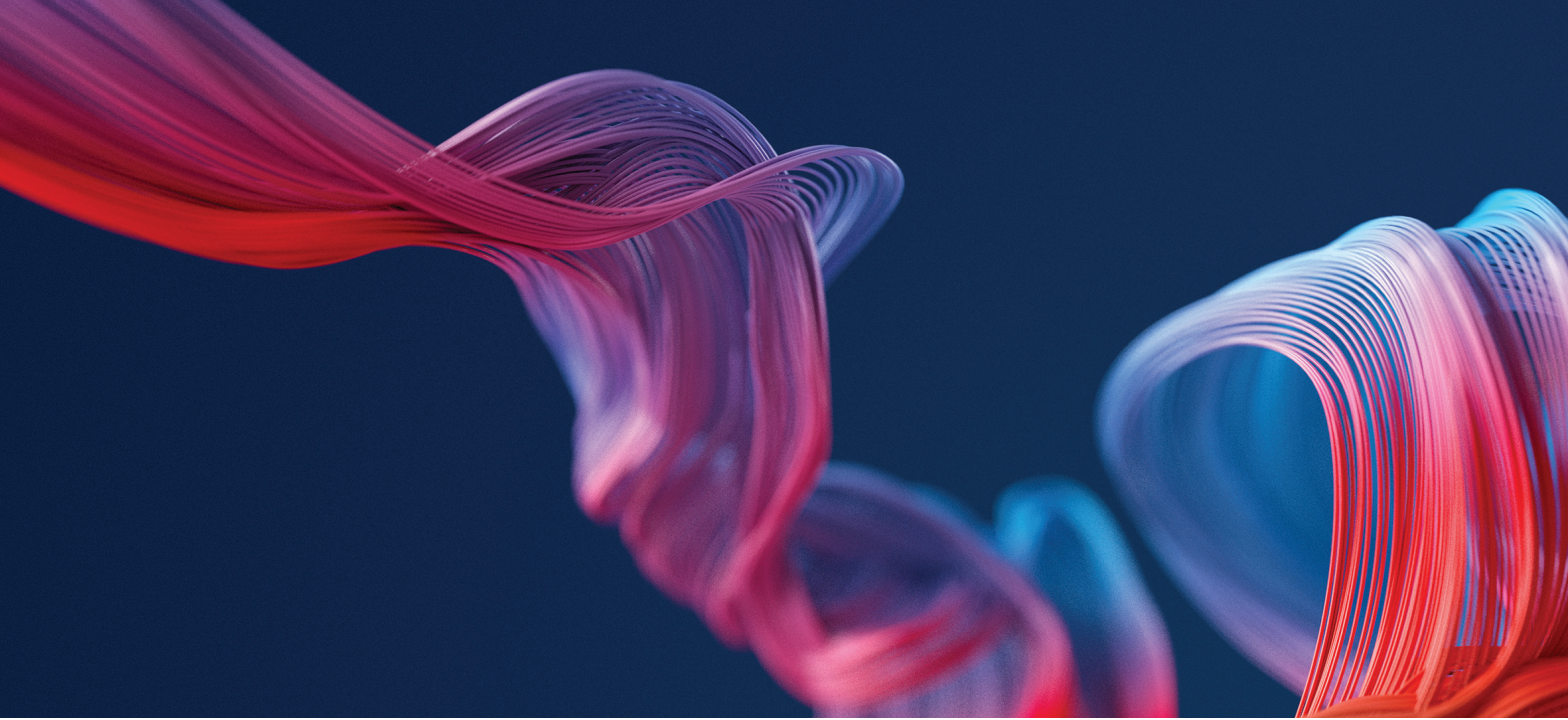

# A brief history of dislocations in ceramics: From Steinsalz to quantum wires

By Xufei Fang

Dislocations in ceramics have enjoyed a long yet underappreciated research history. This brief historical overview and reflection on current challenges provides new insights into using this line defect as a rediscovered tool for engineering functional ceramics.

Dislocations, or one-dimensional line defects, are one of the most fundamental defect types next to zero-dimensional point defects; two-dimensional planar defects, including grain boundaries; and three-dimensional defects, such as precipitates.

As the main carriers for plastic deformation in crystalline solids, dislocations have been extensively studied in metals, which have good deformability. Ceramics, in contrast, are perceived as brittle in the materials science community, exhibiting little or almost no dislocation plasticity at room temperature.

Yet it was fundamental studies on dislocations in ceramics back in the 1950s that significantly contributed to the understanding of dislocations in solids in the early days. In recent years, dislocations in ceramics are seeing a renewed interest owing to the potential advanced functional properties they can help unlock.[1]

In this article, some historical highlights for dislocations in ceramics are gathered, aiming to offer a broader view of the research endeavors as well as the challenges for this reemerging—yet still outlier—topic. This line of historical evolvement hopefully will inspire readers, especially students, to rethink engineering of ceramic materials, which to date has focused primarily on point and planar defects.

## A brief historical overview

While plastic deformation today is most often associated with metals, early-stage observations of bulk plastic deformation was documented in minerals, particularly rock salt (*Steinsalz* in German) in the 1920s.[2] This plastic deformation occurred even at room temperature under tensile loading!

In 1934, G. Taylor, E. Orowan, and M. Polanyi independently conceived the concept of dislocations to explain the mechanism of plastic deformation in solids.[3] In the decades since, roughly three major research waves involving dislocations in ceramics have occurred (Figure 1).

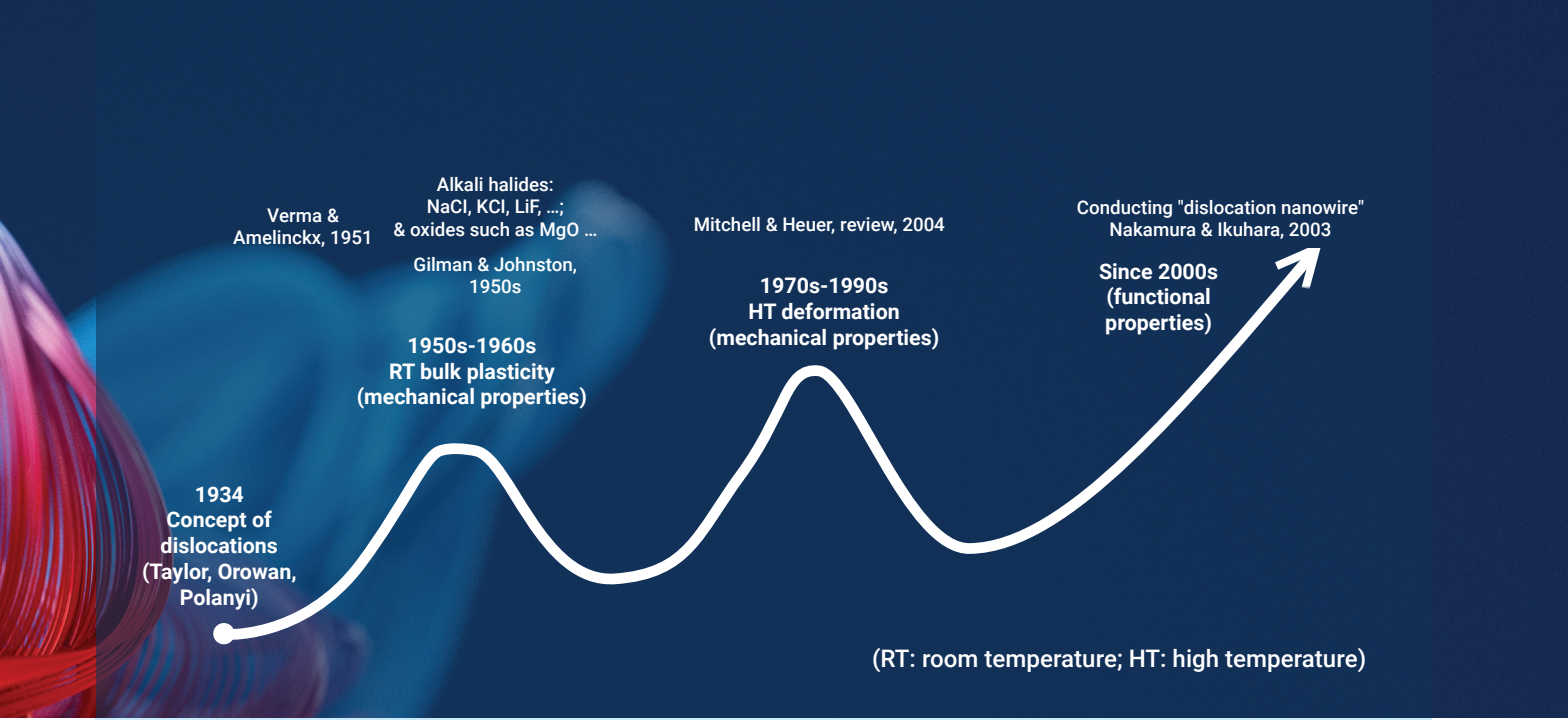

Credit: Xufei Fang

Figure 1. Three major research waves involving dislocations in ceramics have occurred since the concept of dislocations was conceived in 1934. The first two waves focused primarily on dislocation-mediated mechanical properties, with a focus on bulk properties at both room temperature and high temperature. The third wave focuses more on dislocation-mediated functional properties, such as electrical conductivity.

Note: The extensive research on plastic deformation of alkali halides (e.g., rock salt) predating the concept of dislocations is not included in this figure.

Although the focus here is on ceramics, for reasons of completeness and overlapping of the continuous endeavor in dislocations in semiconductors, readers may refer to the short historical review by T. Figielski[4] and the monograph by D. Holt and B. Yakobi.[5]

*First wave: Room-temperature dislocations in ionic crystals*

Probably the first direct observations of dislocations in ceramics were made on silicon carbide featuring spiral growth patterns induced by screw dislocations, which were independently reported by V. Verma and S. Amelinckx in their back-to-back articles in *Nature* in 1951.[6] Shortly after, extensive research on dislocation multiplication, motion, and nucleation in lithium fluoride crystals was carried out by Gilman and Johnston,[7] who used a chemical etching method to directly visualize the dislocation etch pits. This etch pit technique has played a significant role in understanding dislocation behaviors to date.

More studies on dislocations in ionic crystals were later summarized by M. Sprackling in his monograph, *The Plastic Deformation of Simple Ionic Crystals*, in 1976.[8] Meanwhile, for oxides with ionic bonding, J. Pask's group,[9] R. Stokes et al.,[10] and A. Argon and E. Orowan[11] pioneered dislocation research in magnesium oxide in the 1950s and 1960s (primarily at room temperature in bulk deformation, including tensile tests).

Charged dislocations, or linear defects that carry an electric charge, are a unique aspect of ceramics with ionic bonding. This phenomenon caught particular attention in the late 1950s, following the seminal work of J. Eshelby et al.[12]

This interest in charged dislocations extended until about the 1980s, with one of the most comprehensive reviews on charged dislocations by R. Whitworth publishing in 1975,[13] followed by Y. Osip'yan et al. in 1986.[14] The latter focused on the interaction between electric charges and moving dislocations in II-VI group semiconductors, covering the electroplastic effect (dislocation plasticity affected by an electric field) and photoplastic effect (dislocation plasticity influenced by light illumination). Another short review by P. Haasen appeared in 1985, addressing dislocation–point defect interactions in ionic crystals, primarily alkali halides.[15]

*Second wave: High-temperature dislocations in structural ceramics*

Moving on to the 1970s, research on dislocations in ceramics shifted to focus on high-temperature deformation under loading, most likely due to demands in the aerospace industry looking to use high-temperature structural ceramics.[16,17] Representative materials investigated were alumina, zirconia, titania, spinel ($MgAl_2O_4$), and forsterite ($Mg_2SiO_4$).[17] During this period, transmission electron microscopy was extensively

# A brief history of dislocations in ceramics: From Steinsalz to quantum wires

used, providing more direct evidence as well as revealing the complexity of the dislocation structure and configuration.

Realizing the importance of dislocation plasticity in structural ceramics at high temperatures and the fact that in this temperature range "interaction of point defects with dislocations is especially significant," T. Mitchell et al. assembled an overview titled "Interaction between point defects and dislocations in oxides,"[18] after the workshop (with almost the same title) held in the Laboratoire de Physique des Materiaux of the French National Center for Scientific Research in 1978.

*Third wave: Dislocation engineering for functional ceramics*

In 1983, W. Shockley, the founding father of the junction transistor, speculated that dislocations in semiconductor crystals could be used as microwiring,[19] which could be pictured as a conductive tube (along the dislocation core) filled with metallic elements. This concept was experimentally achieved 20 years later by A. Nakamura et al. in the laboratory of Y. Ikuhara.[20] They successfully deformed bulk single-crystal sapphire at 1,400°C and then diffused titanium along the dislocations to achieve an about $10^{13}$ higher conductivity compared to the pristine, insulating sapphire. Based on these proofs-of-concepts, they declared the dislocation-based nanowire design an example of "dislocation technology."[20,21]

About a decade later, S. Szot et al. summarized the use of dislocations in the near-surface regions of bulk oxides, such as strontium titanate and titania, to tune the ceramics' physical and chemical properties.[22] Besides bulk materials, scattered studies also report dislocation-tuned electrical conductivities in oxide thin films, as reviewed by M. Armstrong et al.[23]

Recently, more momentum in the area of dislocation-mediated functional ceramics has been driven by the Ceramics Group led by J. Rödel at Technical University of Darmstadt. The group focused on the mechanical imprinting of dislocations into bulk oxides using high-temperature deformation to achieve enhanced electromechanical properties in ferroelectrics (e.g., barium titanate)[24] and to regulate the electrical conductivity (e.g., in titania)[25] via dislocation self-doping, contrasting the conventional chemical doping strategies using point defects.

The current author, stemming from Rödel's group, has focused on mechanical tailoring of dislocation densities and plastic zone sizes at room temperature across length scales in the model perovskite oxide strontium titanate. His group has also extended the materials systems to potassium tantalate and beyond, with the goal to bring down the temperature required for cost-effective dislocation engineering.[26,27]

For a list of more researchers who have made significant contributions to dislocation science in general, readers may refer to the entertaining webpage by H. Föll.[28]

## Current challenges and the dislocation engineering toolbox

There is still a long way to go to achieve commercialized dislocation-based functional ceramic technology. While dislocations generated via mechanical deformation face the long-standing challenge of crack formation, dislocations produced via novel sintering and other fabrication methods cannot yet achieve precise control of the dislocation structures, except for bicrystal fabrication (which can be rather costly).

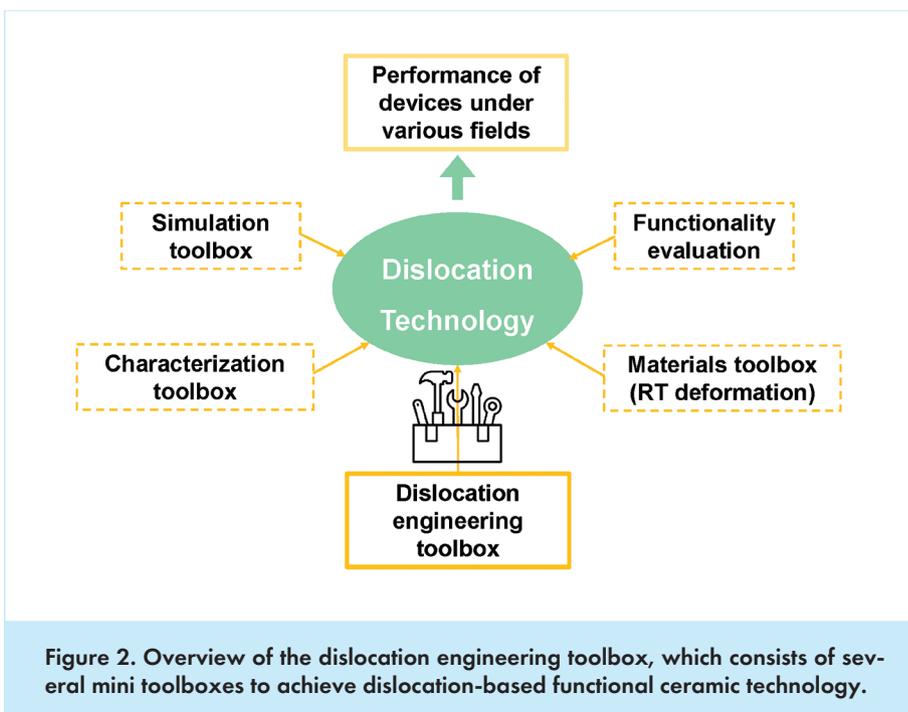

Figure 2. Overview of the dislocation engineering toolbox, which consists of several mini toolboxes to achieve dislocation-based functional ceramic technology.

To this end, the dislocation engineering toolbox can aid development and scaleup of dislocation technologies (Figure 2). This toolbox, which was first presented in Reference 1, consists of several mini toolboxes containing experimental techniques for investigating dislocations in ceramics. It includes resources for not only mechanical deformation across length scales at both room temperature and elevated temperatures, but it also includes novel sintering and other fabrication routes including, for example, flash sintering, thin film growth, bicrystal fabrication, and irradiation.

*Materials toolbox*

Inspired by the discovery of room-temperature bulk plasticity in strontium titanate,[29] potassium niobate,[30] and potassium tantalate,[27] the question has been put forward: Are there more perovskite oxides that are plastically deformable at room temperature?[26] If so, that would help facilitate cost-effective mechanical deformation for dislocation engineering.

The author's group recently reviewed the literature and summarized 44 ceramic compounds that exhibit meso-/macroscale dislocation plasticity at room temperature.[31] This number appears encouraging. Nevertheless, predictions are not available at this stage due to the lack of fundamental understanding concerning dislocation mobility in the majority of ceramics at room temperature, which requires further extensive characterization and simulation.

*Characterization toolbox*

A comprehensive characterization toolbox for dislocation structure ranging from the atomic scale up to the bulk scale is necessary for a complete picture of the fundamental mechanisms of dislocations in ceramics. But currently, the methods in this toolbox are not sufficiently efficient. For instance, capturing reliable 3D reconstruction of the dislocation core structure, which is believed to be most critical for understanding why certain ceramics exhibit good dislocation mobility at room temperature, remains a challenging task. Artifacts can be induced easily from 2D analysis even in state-of-the-art transmission electron microscopy. It is indispensable to obtain accurate experimental information as reliable input for atomistic simulations.

*Simulation toolbox*

The simulation toolbox on dislocations in ceramics may be categorized into two major branches: one on the functional/transport properties and the other on the mechanical/mobility properties. In examining the dislocation–point defect interaction and pipe diffusion, M. Puls and J. Rabier pioneered the simulation of edge dislocations in rock salt and magnesium oxide in the 1980s.[32] In the perovskite oxide strontium titanate, one of the early atomistic simulations to address the role of edge dislocation on the defect chemistry and oxide ion transport was carried out by D. Marrocchelli et al. in 2015.[33]

J. Amodeo, P. Carrez, P. Cordier, and colleagues systematically studied the dislocation motion in magnesium oxide using atomistic simulations, with the aim to provide a more quantitative description of plastic flow in the Earth's mantle.[34,35] The same group also modeled dislocation cores in more perovskites including magnesium silicate, calcium titanate, and strontium titanate. In the meantime, P. Hirel, M. Mrovec, C. Elsässer, and colleagues were examining dislocations in strontium titanate and potassium niobate,[36,37] motivated by the aforementioned discovery of room-temperature bulk plasticity in these perovskite oxides.

The simulation toolbox for dislocations in ceramics is clearly still in its infancy. In particular, the interatomic potential remains the most pressing bottleneck, primarily due to the complex interatomic interactions involving ionic/covalent bonding and extended/dissociated core structures, with the latter heavily relying on accurate experimental observations of core structures, as mentioned above.

Attempts in constructing neural network potentials using density functional theory for large models were made most recently on some ceramics as a potential solution.[38] Scaling it up from atomic scale to address dislocation multiplication, mobility, and hardening at the microscale and eventually connecting to the continuum level remains largely unexplored.

*Functionality evaluation toolbox*

Evaluation of dislocation-tuned functional properties in different structures ranging from thin films to bulk materials are readily available. What is truly critical is the dislocations' long-term stability, which will be directly linked to the performance of future dislocation-based devices. Under such conditions, the complexity will not only arise from the dislocation mesostructure involving different dislocation types (edge, screw, and mixed), kinks, and jogs,[39] but it will also be due to the interactions between dislocations as well as other types of defects, such as point defects and eventually planar defects, e.g., domain walls (single or polycrystalline materials) or grain boundaries (polycrystalline materials).

## Future dislocation technologies

It may seem premature to address dislocation-based functional devices because few attempts to create such devices have been made so far. However, based on the functionalities shown in the research studies described above, we can already posit that these devices may soon find application in a variety of fields, including thermal, electrical, light, magnetic, and mechanical loading technologies.

In addition, a principle noted by W. Shockley at the 1953 March meeting of the American Physical Society may enable the use of dislocations in ceramics for emerging applications.[40] He noted that broken/dangling bonds in crystals with diamond structure may lead to a one-dimensional band of edge states that "may be partially filled, thus causing each dislocation to become a one-dimensional degenerate-electron-gas conductor." This behavior suggests dislocations may be able to act as quantum wires[41,42] or conductive nanowires.[21]

## Concluding remarks

Looking back over the historical development of dislocations in ceramics, there appears to be exciting times ahead for this field. The ground-laying works currently being carried out to overcome the pressing bottlenecks in dislocation engineering in various functional ceramics, irrespective of their brittle nature, may eventually catalyze the realization of dislocation technology for a new generation of functional ceramics.

## Acknowledgments

The author acknowledges the European Research Council (ERC Starting Grant, Project MECERDIS, No. 101076167) for funding the fundamental research. The author would like to thank A. Zelenika and O. Preuß for retrieving and vetting some of the literature from the pre-internet era.

## Disclaimer

Due to the long history and vast amount of literature on this topic scattered in different fields, some novel dislocation-based discoveries and researchers who made significant contributions may not have made it into this review. Readers are invited to contact the author for discussions and exchanges of facts.

## About the author

Xufei Fang leads two research groups, Dislocations in Ceramics and Hydrogen Micromechanics, at Karlsruhe Institute of Technology (KIT), Germany. He is also guest professor at The University of Osaka, Japan. Contact Fang at xufei.fang@kit.edu.

# A brief history of dislocations in ceramics: From Steinsalz to quantum wires